# CORRELATION OF PULSAR RADIO EMISSION SPECTRUM WITH PECULIARITIES OF ELECTRON ACCELERATION IN POLAR GAP

V.M. Kontorovich[1,2], A.B. Flanchik[1]

[1]Institute of Radio Astronomy of the National Academy of Science of Ukraine
4 Krasnoznamionnaya St., Kharkov, 61002, Ukraine
[2]Kharkov Karazin National University
4 Svobody Square, Kharkov, 61077, Ukraine

The analytical expression for frequency of the maximum of radio emission intensity in pulsars with free electron emission from the stellar surface has been found. We have explained the correlation, known from observations, between the high-frequency cut-off and low-frequency turnover in radio emission spectrum of the pulsars. The explanation is based on peculiarities of electron acceleration in the inner polar gap, which we have analyzed.

## 1. INTRODUCTION

The investigations carried out in the Pushchino Radio Astronomy Observatory ASC LPI have shown that pulsars from the 46 objects sample [1-4] have the high-frequency cut-off in the spectrum at frequencies which depends on the pulsar period $P$ [4] (*cf* from "cutoff", the wave over frequency means the experimental value)

$$\tilde{\nu}_{cf} = 1.4 \cdot 10^9 Hz \sqrt{\left(\frac{1\ s}{P}\right)}. \tag{1a}$$

The explanation of this phenomenon, proposed by us in [5-7], connects the cut-off with the change of emission mechanism. In our model the electron emission caused by longitudinal acceleration in the inner gap over the pulsar polar cap, which is responsible for radio emission, stops at frequencies ($B$ is the value of the magnetic field on the pulsar surface)

$$\nu_{cf} = \sqrt{2} \cdot 10^9 Hz \sqrt{\left(\frac{B}{2 \cdot 10^{12} G}\right) \cdot \left(\frac{1\ s}{P}\right)}. \tag{1b}$$

At that it is significant that the accelerating field in the gap (linearly) increases from zero at the stellar surface (see below). In this field, the acceleration of electrons emitted from the pulsar surface also increases from zero and mounts to maximum in the range of sub-relativistic electron velocity. As the velocity approaches to the velocity of light, the electron acceleration decreases, and this emission mechanism terminates. In the narrow cone of directions along the magnetic field line, it is replaced by relativistic emission mechanism [8-10].

As follows from the observations, for some pulsars in the Pushchino sample with the spectrum cut-off, the high-frequency cut-off correlates with the low-frequency turnover. Namely, according to Malofeev and Malov [4], for this sub-sample containing 32 pulsars with known locations of both frequencies, there is the following correlation

$$\tilde{\nu}_{tr} = 0.1 \tilde{\nu}_{cf} . \qquad (2)$$

Here $\tilde{\nu}_{tr}$ (*tr* from turnover) means the frequency, where the intensity of radio emission reaches its maximum and the turnover of the spectrum to lower frequencies begins. For many astrophysical objects, including extragalactic sources, turnover frequency is determined by the dissipative mechanisms or specific plasma dispersion [11]. In our case, these mechanisms can not exist. At the same time, the equation (2) suggests that the low-frequency turnover can be caused by the same acceleration mechanism that leads to the high-frequency cut-off. We will show that this is true.

## 2. ACCELERATION OF ELECTRONS NEAR THE PULSAR SURFACE

Vacuum gap under the magnetosphere of the open field lines was introduced by Ruderman and Sutherland [12]. Strong accelerating electric field almost instantaneously accelerates the electrons, including the electrons torn from the stellar surface by auto-electronic emission, to relativistic velocities raising their gamma-factor up to very high values $\Gamma \sim 10^7$. Momentary burst of acceleration that resembles δ-function by time, after the expanding to Fourier integral, leads to the flat spectrum – the step with the cut-off frequency of the order of the inverse time of acceleration. This frequency falls into the X-ray range, and radio frequencies, respectively, represent a negligible part of the emitted energy. Therefore, others, particularly plasma mechanisms of radio emission in the magnetosphere have been developed. Mainly, the options have been discussed of two-stream instabilities [13], which arise in the plasma with gamma-factor $\Gamma \sim 10^3$, penetrated with the electron beams with $\Gamma \sim 10^7$, as well as some other plasma mechanisms (see reviews [4, 9, 14]). Despite a number of achievements, these mechanisms have not allowed to create a quantitative theory of pulsar radio emission, which would explain in detail all data of observations (see the reasoning in [4, 9]).

The arguments in favor of free emission of electrons from the surface [15] lead to a different picture of distribution of the accelerating field. It is still convenient to talk about the range, in which the accelerating field exists, as about a "gap" that, strictly speaking, cannot be called vacuum one any more [16]. In these models, the longitudinal electric field vanishes at the stellar surface, increases inwards the gap, and vanishes again on its border with the magnetosphere plasma or decreases exponentially in the magnetosphere [17-18]. In this case, the electrons are accelerated at a much longer period than in the model of Ruderman and Sutherland, and with typical parameters of pulsars, the emission caused by the longitudinal acceleration almost completely falls into the radio range.

Since the processes, discussed below, occur at low heights above the stellar surface, it will be sufficient in the future to assume that the accelerating electric field varies from zero at the star surface according to the linear law, disregarding the factors that have led to such dependence.

It is convenient to proceed from the equation for the $\Gamma$-factor, i.e. the expression for the energy E gain of the electron (z is the coordinate orthogonal to the surface, the rest notations are standard)

$$\frac{d\Gamma}{dz} = \frac{eE(z)}{mc^2}, \quad \Gamma \equiv \frac{E}{mc^2} = \frac{1}{\sqrt{1 - V^2/c^2}}. \tag{3}$$

Assuming that the accelerating electric field $E(z)$ is linear near the surface

$$E(z) = E_0 \frac{z}{h}, \quad z \ll h, \tag{4}$$

where $h$ is height of the gap, we find the location-dependent gamma-factor

$$\Gamma(z) = \Gamma_0 + a z^2, \tag{5}$$

where

$$a = \frac{eE_0}{2mc^2 \cdot h}, \tag{6}$$

and $\Gamma_0$ is the gamma-factor of the electron on the pulsar surface. This value is close to 1, but, as we shall see, in this case the presence of electron thermal velocity $V_T \ll c$ is important, and so, for $\Gamma_0$ we will use the expression (see (3))

$$\Gamma_0 \approx 1 + \frac{V_T^2}{2c^2}. \tag{7}$$

More accurate averaging-out of thermal velocities will not be needed here. Expressing the electron velocity through $\Gamma$-factor

$$V(\Gamma) = c\sqrt{1 - \frac{1}{\Gamma^2}}, \tag{8}$$

we find the electron acceleration $w = dV/dt$ [19, 20]:

$$w = \frac{c^2}{\Gamma^3} \frac{d\Gamma}{dz} \tag{9}$$

As it is seen from (3) and (9), the acceleration increases from 0 at z = 0, passes through the maximum (Fig. 1) at $z = z_m$ and tends to zero at the limit $z \gg z_m$. At that [4-7]

$$\Gamma(z_m) = \frac{6}{5}, \quad V(z_m) = c\frac{\sqrt{11}}{6}, \quad z_m = \sqrt{\frac{2mc^2 h}{5eE_0}}. \tag{10}$$

As the accelerating field appears due to rotation of star with angular velocity $\Omega$ in the magnetic field $B$, the estimation on the scale $h$ of the vacuum gap is [17-18]

$$E_0 = \frac{\Omega h}{c} B. \tag{11}$$

The coefficient $a$ in (6) and the coordinate of acceleration maximum $z_m$ do not depend on $h$:

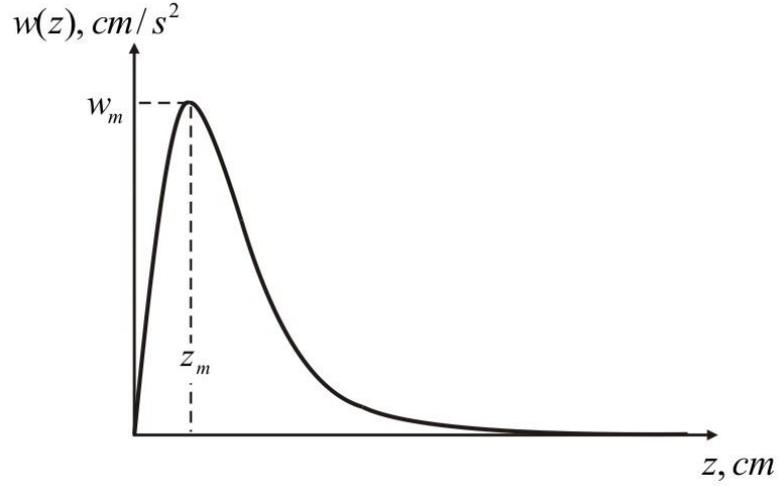

Fig. 1. Dependence of acceleration in the electric field, linearly increasing from zero, on the height above the surface of the star

$$a = \frac{eB}{2mc^2}\frac{\Omega}{c}, \quad z_m = \sqrt{\frac{2mc^3}{5e\Omega B}} \ll h, \tag{12}$$

The latter inequality ($z_m \approx 1$ cm, $h \approx 10^3 - 10^4$ cm for ordinary pulsars with P~1 s and B ~ $10^{12}$ Gs [4, 9]) confirms that in the maximum range we do not go out from the area of applicability of the linear law (4) for accelerating field. In this case, the quadratic term in the Γ-factor (5) is 1/5. The frequency range of the spectrum cut-off $\omega_{cf} = V_{cf}/z_{cf}$, which is determined by the approach of acceleration to zero as electrons reach relativistic velocities [5-7], turns out in the same region $z \ll h$:

$$\Gamma(z_{cf}) = 2, \quad V(z_{cf}) = c\frac{\sqrt{3}}{2}, \quad z_{cf} = \sqrt{5}\, z_m, \tag{13}$$

These relations are described with the condition [5-7]

$$e \int_0^{z_{cf}} E(z)dz = mc^2, \tag{14}$$

which follows from independent physical considerations and determines the location of the spectrum cut-off[1] of radio emission $z_{cf}$ in case of acceleration of the electron in the described conditions. Such behavior of the emission spectrum also follows from numerical analysis of its spectral density (see [5-7] and below).

---

[1] The equality (14) can be refined by detailed comparison with the behavior of spectrum, described by the integral (15).

3. THE TIME OF ELECTRON ACCELERATION TO SUB-RELATIVISTIC VELOCITY

The expression for the emission spectrum is determined by the integral (see Appendix)

$$L_\omega = \int_0^\infty dt \frac{w(z(t))}{(1 - \frac{V(t)}{c}\cos\theta)^2} e^{i\omega\left(t - \frac{z(t)}{c}\cos\theta\right)}, \qquad (15)$$

through which the spectral-angular distribution of the emitted energy is expressed [19, 20]

$$d\varepsilon_{\mathbf{n}\omega} = \frac{e^2}{4\pi^2 c^3}\sin^2\theta |L_\omega|^2 d\omega do, \qquad (16)$$

where $\theta$ is the angle between the electron velocity and the vector of the emitted wave, $do = 2\pi\sin\theta d\theta$. At very low frequencies (15) should allow the expansion into the Taylor series by frequency, since expression (15), as a function of frequency, has no singularity at zero. Since the frequency enters in the exponent in combination $\omega t$, the behavior of acceleration time $t$ is crucial. Let us consider the acceleration time as a function of gamma-factor $t(\Gamma)$. Rewriting

$$dt = \frac{dz}{V} = \frac{1}{V(\Gamma)}\frac{dz}{d\Gamma}d\Gamma, \qquad (17)$$

and taking into account that in this range, according to (5),

$$z(\Gamma) = \sqrt{\frac{\Gamma - \Gamma_0}{a}}, \qquad (18)$$

we get

$$t(\Gamma) = \frac{1}{2c\sqrt{a}}\int_{\Gamma_0}^{\Gamma} d\Gamma \frac{\Gamma}{\sqrt{(\Gamma - \Gamma_0)(\Gamma^2 - 1)}}. \qquad (19)$$

First we will consider the natural extreme case of zero electron velocities on the stellar surface: $\Gamma_0 \to 1$. Then

$$2c\sqrt{a}\, t(\Gamma) \to \int_{\Gamma_0 \to 1}^{\Gamma} d\Gamma \frac{\Gamma}{(\Gamma - 1)\sqrt{\Gamma + 1}}. \qquad (20)$$

The integral in the right part of (20) diverges logarithmically at the lower limit, which we will illustrate isolating in the integrand expression the special part[2] - contribution from the vicinity of the lower limit:

---

[2] When $\Gamma_0 \to 1$, $1/(\Gamma - 1)$ in the integrand is a sharp function of the integration variable, which reaches its maximum at the lower limit. Therefore, the main contribution into the integral is made by its neighborhood. We substitute the value of the variable at the lower limit $\Gamma_0 = 1$ into the slowly varying factors. (This corresponds to the Laplace method, but because of insufficiently rapid decay of the sharp function, the numerical coefficient in the logarithm requires some correction.)

$$\int_{\Gamma_0}^{\Gamma} d\Gamma \frac{\Gamma}{(\Gamma-1)\sqrt{\Gamma+1}} \approx \frac{1}{\sqrt{2}} \int_{\Gamma_0}^{\Gamma} \frac{d\Gamma}{\Gamma-1} = \frac{1}{\sqrt{2}} \ln \frac{1}{\Gamma_0 - 1}. \qquad (21)$$

This means that in the accelerating field, linearly increasing from zero, the electron with zero initial velocity requires logarithmically infinite time to reach a terminal velocity. Physically, this condition is quite understandable. With zero initial velocity and zero acceleration at the stellar surface, the electron will not move at all and will remain on the surface for indefinitely long time. Therefore, the presence of initial (thermal) velocities of electrons becomes crucial. As we will show, this divergence is the cause of law-frequency turnover.

If to substitute the lower limit $\Gamma_0 \neq 1$ into (20), we get (leaving at first only the contribution from the lower limit and, accordingly, replacing $\Gamma/\sqrt{\Gamma+1} \to 1/\sqrt{2}$ in the integrand) the value of characteristic acceleration (pre-acceleration) time $\bar{t}$ of the electron

$$2c\sqrt{a}\,\bar{t} = \frac{1}{\sqrt{2}} \ln \frac{2c^2}{V_T^2}. \qquad (22)$$

Further refinements are connected with the numerical coefficient of order 1 in the logarithm. So, retaining $\Gamma_0$ not only in the limit, but also in the integrand (19), instead of (22) we get the following contribution from the lower limit [22]

$$2c\sqrt{a}\,\bar{t} = \frac{1}{\sqrt{2}} \ln \frac{4c^2}{V_T^2}. \qquad (23)$$

Thus, the logarithmic behavior of the time of acceleration (Fig. 2) from thermal to the logarithmically large epithermal velocity leads to the logarithmically large time step $\bar{t}$, which the electron has to overcome before it enters the acceleration regime. This behavior leads to low-frequency turnover in the spectrum (see Fig. 3).

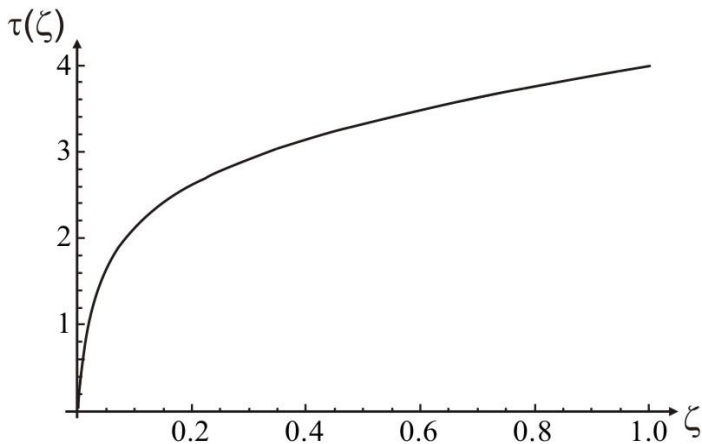

Fig. 2. Dependence of the dimensionless acceleration time on the dimensionless distance $\zeta \equiv z/z_{cf}$ in the accelerating field $E(z) = z \cdot \Omega B/c$ for $\Gamma_0^2 - 1 = 10^{-4}$. There are two significantly different time scales for small and large distances.

Expressing *a* through the frequency of the high-frequency cut-off $\omega_{cf}$ according to

$$\omega_{cf} = \frac{2\pi c}{z_{cf}}, \quad z_{cf} = \frac{1}{\sqrt{a}}, \quad \omega_{cf} = 2\pi c\sqrt{a}, \qquad (24)$$

get the temporary step

$$\bar{t} = \frac{\pi}{\sqrt{2}} \frac{1}{\omega_{cf}} \ln \frac{4c^2}{V_T^2}. \qquad (25)$$

## 4. EVALUATION OF SPECTRUM TURNOVER AND EMISSION MAXIMUM FREQUENCIES

The possibility expansion of function $L_\omega$ into the series by low frequencies is determined by the condition

$$\omega \bar{t} \ll 1, \qquad (26)$$

when the exponent can be expanded into the series and the logarithmic singularity of the integral (15) has no effect on its behavior. In this frequency range, the emission intensity increases with the increase of frequency.

According to the expression for $\bar{t}$ obtained from (25), from (26) we get:

$$\omega \ll \frac{\omega_{cf}}{\frac{\pi}{\sqrt{2}} \ln \frac{4c^2}{V_T^2}}. \qquad (27)$$

At the same time, the emission spectrum of pulsar at higher frequencies, which, according to the considered mechanism, increases to low frequencies under power law [5-7], will be limited by the frequency determined from the inequality (27). For the maximum spectrum frequency $\omega_{max} = \omega_{tr}$ we can assume the estimate (to the dimensionless factor of order 1) determined by the right side of this inequality, bringing in there for convenience an additional factor 1/2 :

$$\omega_{tr} = \frac{\omega_{cf}}{\frac{\pi}{2\sqrt{2}} \ln \frac{4c^2}{V_T^2}}. \qquad (28)$$

Such definition $\omega_{tr}$ sufficiently accurately corresponds to correlation (2) between the cut-off frequency and maximum spectrum frequency [4]. Indeed, according to [23], temperature of the pulsar surface reaches one million degrees. Then $\frac{4c^2}{V_T^2} \approx 10^4$, $\log \frac{4c^2}{V_T^2} \approx 4$, $\ln \frac{4c^2}{V_T^2} \approx 9.2$, which gives the required estimation (2)

$$\omega_{tr} \approx 0.1\, \omega_{cf}. \qquad (29)$$

The temperature dependence of the pulsar surface contained in (28), will probably allow to use (28) as a kind of thermometer to measure the electron temperature of the surface.

## 5. BEHAVIOR OF THE SPECTRUM AT LOW FREQUENCIES

Acceleration the electron to relativistic energies, as seen in Fig. 2, is a two-stage process. Initially, at low (thermal) initial velocity and the acceleration slowly increasing from zero, the electron, for a logarithmically long period, moves with velocity of order of thermal one and slowly starts to enter in the acceleration regime. This interval (of the order of the inverse frequency of spectrum cut-off, divided by a large logarithmic factor (22)) also defines the maximum spectrum frequency (28).

During the next, much shorter period of time, the electron quickly gains its velocity from epithermal to relativistic ones (Fig. 2). The duration of this part of the trajectory has the order of the inverse frequency of spectrum cut-off. The movement in this region is determined by acceleration in the longitudinal field $E_\parallel$.

At higher frequencies (due to rapid oscillations under the Fourier integral sign), the emission in the wide range stops. This corresponds to transition to relativistic movement with vanishingly small acceleration. With large gamma factors, the emission remains essential only in a narrow aberration cone along the direction of the movement (direction of magnetic field line). The low-frequency tail of this emission falls into the radio range too, along with the low-frequency tail of the curvature radiation. We do not consider it here (see [6]).

For the analysis of the mentioned properties of the emission in the pre-acceleration region, corresponding to low frequencies, it is convenient to proceed from the expression for the emitted energy in the form of

$$d\varepsilon_{\mathbf{n}\omega} = \frac{e^2}{4\pi^2 c}\sin^2\theta \,|\frac{V_T}{c}+iD_\omega|^2 \, d\omega do, \quad D_\omega = \frac{\omega}{c}\int_0^\infty V(t)\exp\left\{i\omega\left(t-\frac{z}{c}\cos\theta\right)\right\}dt, \qquad (30)$$

which is obtained by integration by parts of the expression (15) (see [20] and the Appendix). The summand $V_T/c$ originates from the contribution of the lower limit in the integrated term. Due to its contribution, as it should be according to [19, 20], the frequency dependence disappears at the lowest frequencies. We divide the interval of integration into several parts: pre-acceleration part from 0 to $\bar{t}$ (25), acceleration part in the field from $\bar{t}$ to sub-relativistic velocities, determined by the high-frequency cut-off of the spectrum $\bar{\bar{t}} = 2\pi/\omega_{cf}$, and the part of relativistic movement from $\bar{\bar{t}}$ to $\infty$ (see the Appendix). Since we are interested in low-frequency emission outside the cone of relativistic aberration, we can omit the second term in the exponent (30) and consider the non-relativistic movement. In the first, longest part of the route, using the mean value theorem, we have the estimate

$$D_\omega \approx \frac{V_T}{c} \cdot \ln\frac{2c^2}{V_T^2} \cdot (e^{i\omega\bar{t}} - 1) \to i\omega\bar{t} \cdot \frac{V_T}{c} \cdot \ln\frac{2c^2}{V_T^2} \quad \text{for } \omega\bar{t} \ll 1. \qquad (31)$$

So, the intensity increases with increasing frequency according to the square law (Fig. 3). In the acceleration part in the frequency range $1/\bar{t} \ll \omega \ll \omega_{cf}$, the electron velocity can be taken as $c/2$, which gives

$$D_\omega \approx \frac{1}{2} \cdot e^{i\omega\bar{t}} (e^{i\omega/\omega_{cf}} - 1). \qquad (32)$$

Another estimate can be obtained if to assume that on this segment of the route the velocity changes faster than the exponent. Therefore, we can take out $e^{i\omega\bar{t}}$ from the integral sign

$$D_\omega = \frac{\omega}{c} e^{i\omega\bar{t}} \int_{\bar{t}}^{\omega_{cf}^{-1}} V(t)dt = \frac{\omega}{c} e^{i\omega\bar{t}} \int_{z(\bar{t})}^{z_{cf}} dz \approx \frac{\omega}{c} z_{cf} e^{i\omega\bar{t}}. \qquad (33)$$

Since the frequency range is small, for estimation the factor $\omega$ can be replaced by the factor $\omega_{cf}$. We observe the modulation of $D_\omega$ by frequency with period $1/\bar{t}$, which, in this rough description, may disappear when substituted into the expression for the intensity. In more accurate analysis the oscillations with maximum spectrum frequency (period $\bar{t}$) should remain. These oscillations are clearly visible in the numerical calculation of the integral (30) (Fig. 3).

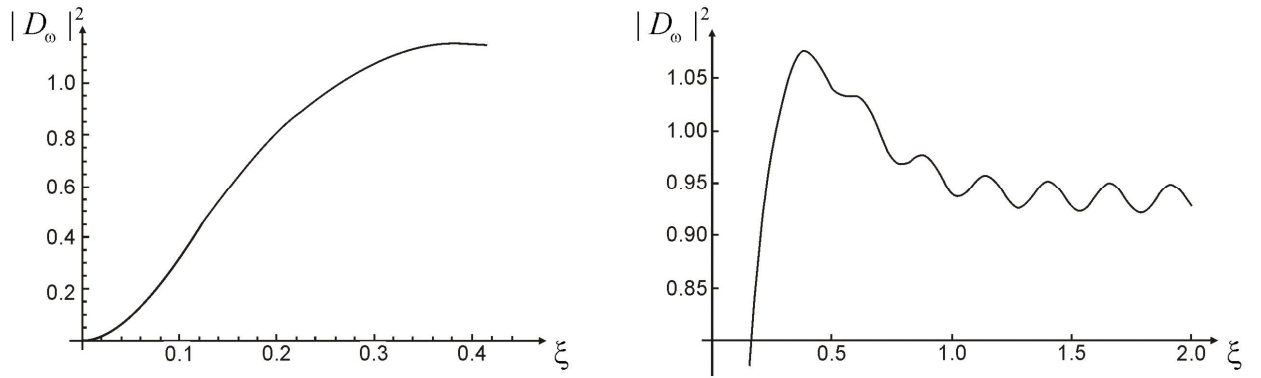

Fig.3. Low-frequency spectrum turnover at electron acceleration in the longitudinal electric field, linearly increasing from zero at the stellar surface. The scale of the dimensionless frequency $\xi = \omega / \omega_{cf}$ is different in the left and right parts of the figure.

Apparently, the oscillations by frequency do not contradict the observational data for some pulsars (see the examples of spectra in [2-3]), but here special analysis is required, especially as the spectra of pulsars have already are averaged for eliminating the scintillations. At higher frequencies, the spectrum cut-off should occur. Integral in the form (30), however, is ill-suited for numerical analy-sis in the high-frequency range because of poor convergence. The initial expression (15) clearly shows the high-frequency cut-off of the spectrum [5-7].

Full analysis of behavior of the emission spectrum at low frequencies in the emission mechanism under consideration requires a discussion about of coherent radio emission at high frequencies and its "switching-off" at the transition to low frequencies, which has been never done yet. Of necessity, it must be based on the model concepts of the structure of the emitting region, more detailed than in [5-7] and in the given paper.

## CONCLUSIONS

This paper presents the theoretical explanation of correlation between the high-frequency cut-off and low-frequency turnover of the pulsar spectrum, which was observed in the Pushchino sample [5-7]. We have derived the expression (28) for the frequency of turnover (that is, simultaneously, is the maximum frequency in the pulsar spectrum), which allows the independent observational verification, since the maximum frequency depends, except for the pulsar period, on the magnetic field and the star surface temperature. Besides, studying the dependence of frequency $\omega_{tr}$ on the physical parameters for the pulsars for which not observed cut-offs, may shed light upon the physical nature of emission, forming their spectra.

Apparently, various emission mechanisms can coexist in such complex systems as pulsars. The considered emission in the inner gap does not exclude the possibility of emission due to other mechanisms at other heights. Prevalence of this or that mechanism is defined by relations of physical parameters.

It is noteworthy that, except for thermal velocities, on which the location of the cut-off frequency (maximum of the spectrum) essentially depends in this model, an alternative option arises from consideration of fluctuations of the accelerating field on the stellar surface. Such consideration (as well as consideration of temperature fluctuations) should result in heterogeneity of acceleration process and formation of electron streams ejected from the stellar surface. This could essentially affect the overall picture of physical processes in the inner gap, and, consequently, the properties and characteristics of radio emission of pulsars.

***

## APPENDIX

1. Transform the expression for the spectral-angular distribution of energy emitted by the charge

$$d\varepsilon_{\mathbf{n}\omega} = \frac{e^2}{4\pi^2 c} |\mathbf{L}_{\mathbf{n}\omega}|^2 do, \quad \mathbf{L}_{\mathbf{n}\omega} = \int_{-\infty}^{\infty} \frac{[\mathbf{n}[\mathbf{n} - \boldsymbol{\beta}, \dot{\boldsymbol{\beta}}]]}{(1 - \mathbf{n}\boldsymbol{\beta})^2} e^{i\omega(t - \frac{\mathbf{n}\mathbf{r}(t)}{c})} dt, \quad \boldsymbol{\beta} = \frac{\mathbf{V}}{c} \tag{A1}$$

into the form suitable for the analysis of low-frequency emission of the charge in longitudinal electric field, linearly increasing from zero. For this, take into account [20] that

$$\frac{[\mathbf{n}[\mathbf{n}-\boldsymbol{\beta},\dot{\boldsymbol{\beta}}]]}{(1-\mathbf{n}\boldsymbol{\beta})^2} = \frac{d}{dt}\left(\frac{[\mathbf{n}[\mathbf{n},\boldsymbol{\beta}]]}{1-\mathbf{n}\boldsymbol{\beta}}\right). \qquad (A.2)$$

At the longitudinal acceleration under consideration $\boldsymbol{\beta} = \beta \mathbf{e}_z$, $\dot{\boldsymbol{\beta}} = \dot{\beta}\mathbf{e}_z$, and the vector factor can be taken out from the integral by time. Given that the $\beta(t) = 0$ at $t < 0$, we get

$$d\varepsilon_{\mathbf{n}\omega} = \frac{e^2 \sin^2\vartheta}{4\pi^2 c}|L_\omega(\vartheta)|^2 do, \; L_\omega(\vartheta) = \int_0^\infty \frac{\dot\beta}{(1-\beta\cos\vartheta)^2} e^{i\omega(t-\frac{z(t)}{c}\cos\vartheta)} dt. \qquad (A.3)$$

Let us integrate the expression for $L_\omega$ by parts taking into account $\dfrac{\dot\beta}{(1-\beta\cos\vartheta)^2} = \dfrac{d}{dt}\left(\dfrac{\beta}{1-\beta\cos\vartheta}\right)$:

$$L_\omega(\vartheta) = \lim_{\substack{T\to\infty \\ \alpha\to+0}} \int_0^T \frac{\dot\beta}{(1-\beta\cos\vartheta)^2} e^{i\omega(t-\frac{z(t)}{c}\cos\vartheta)} dt = \lim_{\substack{T\to\infty \\ \alpha\to+0}}\left\{\beta e^{i(\omega+i\alpha)(t-\frac{z(t)}{c}\cos\vartheta)}\Big|_0^T - iD_\omega(\vartheta)\right\}, \qquad (A.4)$$

$$D_\omega(\vartheta) = \omega \lim_{\substack{T\to\infty \\ \alpha\to+0}} \int_0^T \beta e^{i(\omega+i\alpha)(t-\frac{z(t)}{c}\cos\vartheta)} dt. \qquad (A.5)$$

In contrast to the results adduced in [20], in our case the contribution from the term outside the integral is significant. Since the electron velocity tends to the constant limit $\beta \to 1$, the integral (A5) converges poorly, and we have to introduced regularization in the form of vanishingly small imaginary part of the frequency $\omega \to \omega + i\alpha, \; \alpha \to +0$. Then the contribution from the upper limit in the integrated term vanishes, and we come to formula (30) in the main text.

2. Consider the integral

$$ct(z) = \int_0^z dx \frac{\Gamma_0 + ax^2}{\sqrt{(\Gamma_0 + ax^2)^2 - 1}}, \qquad (A.6)$$

describing the dependence of time $t(z)$ on the electron coordinate during acceleration from the thermal to relativistic velocities in the electric field, linearly increasing from zero. We have used the dependence of gamma-factor on the coordinate (5), where the term $az^2$ is limited only by the condition $z \ll h$ and may be of order or more than 1. In particular, at the point of the spectrum cut-off [5-7], $\Gamma_{cf} = 2$ and $az_{cf}^2 = 1$. Let us pass into the dimensionless variables $\tau \equiv \omega_{cf} t/2\pi$ and $\zeta \equiv z/z_{cf}$. Coefficient $a$ in (5) is $a = z_{cf}^{-2}$, and (5) becomes $\Gamma(z) = \Gamma_0 + \zeta^2$. Instead of (A.6), we have

$$\tau(\zeta) = \int_0^\zeta d\xi \frac{\Gamma_0 + \xi^2}{\sqrt{(\Gamma_0 + \xi^2)^2 - 1}}. \qquad (A.7)$$

It is obvious that at zero "initial" velocity in the considered field, the electron will not accelerate at all. At low thermal velocity $V_T \ll c$ on the smallest heights, its movement will be very slow, and it will need very long time to gain velocity, significantly exceeding the thermal one. This part of the path (where we assumed $\Gamma_0 + 1 \to 2$ and omitted the terms with $\xi^2$ in the integrand) corresponds to the asymptotics of the integral (A.7) equal to

$$\tau(\zeta) = \frac{\zeta}{2\sqrt{\Gamma_0 - 1}}, \quad \zeta << \sqrt{\Gamma_0^2 - 1}. \tag{A.8}$$

In the region, where the velocity starts to differ substantially from the thermal one, a logarithmic asymptotic arises

$$\tau(\zeta) = \frac{1}{\sqrt{2}} \int_{\Gamma_0 - 1}^{\zeta_*} \frac{d\xi}{\xi}, \quad \Gamma_0^2 - 1 << \zeta_* << 1, \tag{A.9}$$

with the contribution from the lower limit determining the time scale of the initial acceleration region $\tau = \bar{t}$ in accordance with the expression (25) of the main text. It corresponds to the expansion of the integrand (A.7) by $(\Gamma_0^2 - 1)/\xi^2 << 1$ with taking into account $\xi^2 << 1$.